\documentclass[utf8]{frontiersSCNS} 

\usepackage{url,hyperref}
\usepackage[onehalfspacing]{setspace}
\usepackage{graphicx}
\usepackage{url}                                                              
\usepackage[]{natbib}     
\usepackage{epsfig}
\usepackage{color}
\usepackage{xspace}
\usepackage{hyperref}

\definecolor{DarkGreen}{rgb}{0.0,0.4,0.0}  

\newcommand{\aap}{    {\it Astron. Astrophys.}}

\newcommand{\apj}{    {\it Astrophys. J.}}
\newcommand{\apjl}{   {\it Astrophys. J. Lett.}}

\newcommand{\apss}{   {\it Astrophys. Space Sci.}}

\newcommand{\nat}{    {\it Nature}}

\newcommand{\pasj}{   {\it Pub. Astron. Soc. Japan}}
\newcommand{\solphys}{{\it Solar Phys.}}
\newcommand{\ssr}{    {\it Space Sci. Rev.}}

\def\keyFont{\fontsize{8}{11}\helveticabold }
\def\firstAuthorLast{Awasthi \& Liu} 
\def\Authors{Arun Kumar Awasthi\,$^{1}$ and Rui Liu\,$^{1}$}
\def\Address{$^{1}$CAS Key Laboratory of Geospace Environment, Department of Geophysics and Planetary Sciences, University of Science and Technology of China, Hefei 230026, China }
\def\corrAuthor{Arun Kumar Awasthi}
\def\corrEmail{arun.awasthi.87@gmail.com}

\begin{document}
\onecolumn
\firstpage{1}

\title[Kinked flux rope magnetic configuration of a prominence bubble]{Mass motion in a prominence bubble revealing a kinked flux rope configuration}

\author[\firstAuthorLast ]{\Authors} 
\address{} 
\correspondance{} 

\extraAuth{Rui Liu \\ rliu@ustc.edu.cn}

\maketitle

\begin{abstract}
\section{Prominence bubbles are cavities rising into quiescent prominences from below. The bubble-prominence interface is often the active location for the formation of plumes, which flow turbulently into quiescent prominences. Not only the origin of prominence bubbles is poorly understood, but most of their physical characteristics are still largely unknown. Here, we investigate the dynamical properties of a bubble, which is observed since its early emergence beneath the spine of a quiescent prominence on 20 October 2017 in the H$\alpha$ line-center and in $\pm$0.4~\AA~line-wing wavelengths by the 1-m New Vacuum Solar Telescope. We report the prominence bubble to be exhibiting a disparate morphology in the H$\alpha$ line-center compared to its line-wings' images, indicating a complex pattern of mass motion along the line-of-sight. Combining Doppler maps with flow maps in the plane of sky derived from a Nonlinear Affine Velocity Estimator, we obtained a comprehensive picture of mass motions revealing a counter-clockwise rotation inside the bubble; with blue-shifted material flowing upward and red-shifted material flowing downward. This sequence of mass motions is interpreted to be either outlining a kinked flux rope configuration of the prominence bubble or providing observational evidence of the internal kink instability in the prominence plasma.}
\tiny
 \keyFont{ \section{Keywords:} Quiescent Prominences, Prominence: Bubble, Prominence: Magnetic field, Prominence: Instability, Kinked flux rope} 
\end{abstract}

\section{Introduction}

It is crucial to understand the dynamical characteristics and magnetic field configuration of the solar prominences primarily due to their association with the solar eruptions. Prominences are believed to be a visible manifestation of the cool material suspended in the corona on the dips of the nearly horizontal magnetic field lines spanned across the polarity inversion line \citep{Leroy1984, Aulanier2002, Ariste2015}. Although a long-term observational history of solar filaments enabled the general characterization of its formation and evolution \citep{Martin1994, Mackay2010, Parenti2014}, high-resolution observations unveiled obscure features, for example, the prominence bubbles. Bubbles are observed as a void region located just above the spicule height emerging underneath the quiescent prominences \citep{Stellmacher1973, Berger2012}. Bubbles are known to be the active locations for the formation of `plumes' which are a probable source of mass supply into the prominence \citep{Berger2008} countering the observed drainage of the prominence material due to the gravitation pull. Therefore, it is crucial to determine the physical mechanisms responsible for the formation, uprise, and expansion of the bubbles.

It is unclear what is inside the prominence bubble. Is it a void region or filled with low-density cool plasma? The earliest observation of the prominence bubble in Ca 8542~\AA~spectra \citep{Stellmacher1973} revealed the absence of line emission in the bubble to be due to its absorption by the cool plasma, and not due to the ``off-band effect'' of the filter. \citet{Heinzel2008} compared the EUV and X-ray intensities observed from prominence with the bubble, determining the opacity of the bubble (and hence the hydrogen column density) to be approximately one-sixth of that of the prominence. Similarly, \citet{Labrosse2011} obtained the intensity of the coronal Fe~\textrm{xii}~line in the bubble to be larger than that in the prominence, however lower than the corona. They speculated the absorption to be due to the optically thin prominence plasma which, however, is not clearly visible in the H$\alpha$ images. \citet{Berger2012} determined the temperature of the plasma inside the bubble to be ranging between 2.5-12$\times$10$^5$ kelvin, and that it is 25-120 times hotter than the surrounding prominence material. Of particular interest was the observation of a hot rising structure (logT$\approx$6.0) within a prominence bubble investigated in \citet{Berger2012}, which was argued to play a crucial role in the formation and expansion of the bubble through pushing the cooler prominence material upwards. They further inferred ``magneto-thermal convection" process to be responsible for the expansion of the prominence bubble. Similarly, \citet{Shen2015} also found higher temperature inside the bubble ($<T>$=6.83MK) compared to the surrounding prominence material  ($<T>$=5.53MK). On the other hand, \citet{Dudik2012} rejected the presence of any kind of hot material inside the bubble based on the investigation of a prominence bubble in the EUV 193~\AA~wavelength. Further, the emission inside the prominence bubble and the prominence cavity region was found to be of similar magnitude. In agreement to this, \citet{Gunar2014} interpreted the apparent brightening in the bubble region (particularly in the 171~\AA~images) to be due to the material corresponding to the prominence-corona-transition-region (PCTR) in the foreground or the background of the bubble based on the observational evidence that the bubble appeared as a void region in H$\alpha$ images but not distinctly separable in the contemporaneous optically thick EUV 304~\AA~ images. Therefore, it is evident that the consensus on the composition of the bubble interior is yet to be reached.

Highly structured ambient and background magnetic field on the top of low emitting bubbles makes it very difficult to determine the magnetic field strength and configuration of the prominence bubble. \citet{Dudik2012} modeled prominence bubble through the inclusion of an emerging parasitic bipole beneath the prominence where the arcade field lines of the bipole correspond to the boundary of the bubble. They further argued that bubbles are devoid of any material and just the ``gaps or windows" in the prominence due to the absence of dips in the bubble field lines. Moreover, the reconnection between the arcade field lines of the bubble and that of the overlying prominence may explain the generation of plumes. \citet{Shen2015} interpreted the enhanced temperature inside the investigated bubble using the aforesaid scheme of reconnection. The magnetic field strength of two different prominence bubbles estimated using the \emph{THEMIS/MTR} polarimetric observations revealed a higher magnetic field inside the bubble compared to the prominence \citep{Levens2016}, which is indicative of emerging magnetic flux at the location of the prominence bubble. Observational evidences of flux emergence beneath the prominence can be found in \citet{Chae2001}. Based on these observations, the role of Lorentz force has been proposed to explain the emergence and uprise of the prominence bubbles.

Quiescent prominences exhibit irregular motion persistently over the entire structure which is generally attributed to the fundamental plasma instabilities. \citet{Ryutova2010} investigated several cases of prominence with bubbles and plumes and suggested the presence of both the Kelvin--Helmholtz (K--H) and Rayleigh--Taylor (R--T) instabilities. K--H instability is attributed to driving the ripples (perturbations) at the bubble boundary to form a single large plume whereas self-similar multiple plumes are suggested to be due to the R--T instability. Further, they characterized the bubble to be a ``growing coronal cavity'' underneath the prominence and suggested the screw-pinch instability \citep{Sakurai1976} to be the formation mechanism of the bubble. K-H instability is one among the most commonly observed instabilities \citep{Zhelyazkov2016} and occurs at the surface of discontinuity of the two fluids which propagate with different speeds, however, possess sufficient enough shear so as to overcome the surface tension force. \citet{Berger2017} attributed the coupled KH--RT instability to be responsible for the development and growth of ripples at the boundary of a prominence bubble as they were located at the density inversion layer. Similarly, \citet{Mishra2019} found the magnetic R-T (MRT) instability to drive regular formation and development of plumes originating from small-scale cavities developed within the prominence whereas the collapse of a plume was attributed to K-H instability. Thus, probing the dynamical behavior of the prominence material leads to the identification of associated plasma instabilities, and in turn, offers insights into the physics of formation and stability of the quiescent prominence.

Therefore, it is evident that the physics of the formation and evolution of prominence bubbles is still debated. Thanks to high spatial-resolution H$\alpha$ images of a prominence recorded by New Vacuum Solar Telescope (NVST), we distinctly characterize the mass motions within a prominence bubble (section~\ref{sec:results}) in this work. A crucial finding of our analysis is the presence of disparate mass distribution within the bubble in the co-temporal H$\alpha$ line-center and in line-wing images. An interesting feature in the EUV observations of the prominence is the presence of a ``bright compact region'' within the bubble. The morphological and thermodynamical evolution of the blob is made to discuss its origin in the context of bubble or from PCTR. Finally, Doppler analysis from the H$\alpha$ line-wing observations is employed to infer the magnetic skeleton of the prominence bubble in section~\ref{sec:conc}.

\section{Instruments and Data} \label{sec:obs}
\begin{figure}[!htbp]
\centering
\includegraphics[width=0.7\textwidth]{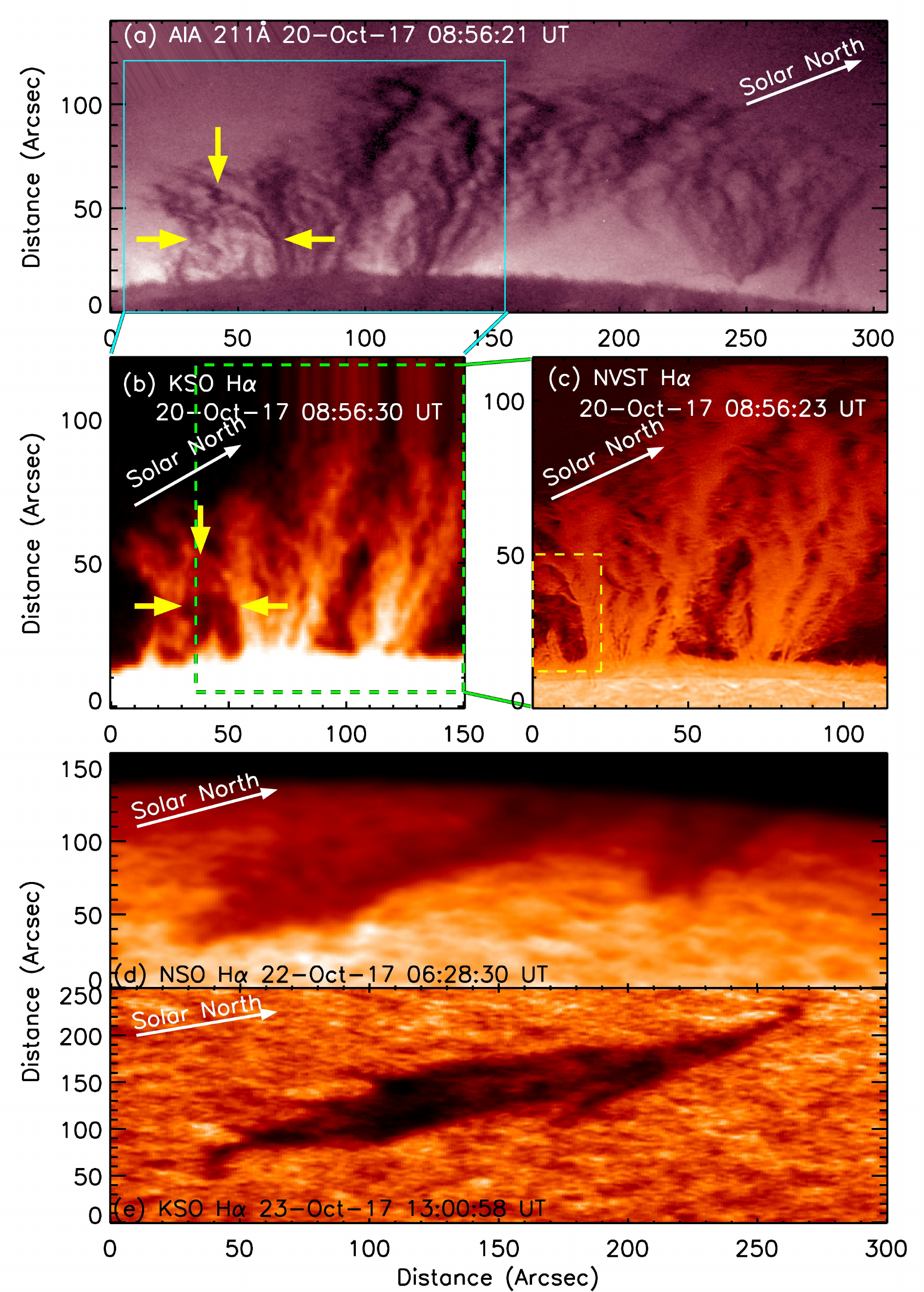}
\caption{Multi-wavelength overview of the quiescent prominence, located at the northeast limb of the solar disk on October 20, 2017 (N24E87). All the images have been rotated clockwise by 65$^\circ$ for presenting an upright view of the prominence. (a) AIA 211~{\AA} image showing the entire prominence spanning across approximately 250" on the limb. The Cyan rectangle represents the field of view of the H$\alpha$ image acquired by the Kanzelhoehe Solar Observatory (KSO), shown in (b). The green square region in (b) indicates the field-of-view of the H$\alpha$ image recorded by NVST, shown in panel (c). Prominence bubble, investigated in this paper is outlined in yellow in panels a--c. (d)--(e) Prominence as observed on 22 and 23 October 2017 from the \emph{National Solar Observatory (NSO)} and \emph{KSO}, respectively. It is clear that the prominence morphology has not altered significantly.}\label{fig:1}
\end{figure}

In order to investigate the mass motion in a quiescent prominence, we primarily use the images acquired by a ground-based 1-m New Vacuum Solar Telescope (NVST; \citet{Liu2014b}) in the H$\alpha$ line center and in $\pm$0.4~{\AA} wings, during 07:27 -- 09:28 UT. NVST raw data-set has been further subjected to the alignment as well as speckle reconstruction, resulting in the pixel scale and the temporal cadence of the final H$\alpha$ images to be $0''.136$ and 28 s, respectively. While the NVST field-of-view could only capture the southern section of the prominence (Figure~\ref{fig:1}a--c), the full context of the prominence has been obtained using full-disk extreme ultra-violet (EUV) images in the 211~{\AA}, 171~{\AA}, and 193~{\AA} wavelengths, recorded by the Atmospheric Imaging Assembly \citep[AIA;][]{Lemen2012} onboard \emph{Solar Dynamics Observatory} \citep[SDO;][]{Pesnell2012} with a pixel scale of $0''.6$ and a temporal cadence of 12~s. Full-disk H$\alpha$ images acquired by the \emph{Kanzelh$\ddot{o}$he Solar Observatory (KSO)}  and \emph{Global Oscillation Network Group (GONG)} with the pixel scale of $1''$ have also been utilized to study the long-term evolution of the prominence.

\section{Observational Results} \label{sec:results}
We investigate the mass motion in a quiescent prominence of October 20, 2017, located at the north-east limb of the solar disk (N24E87). In particular, dynamical characteristics intrinsic to a prominence bubble (marked by yellow arrows in the panels a--c of Figure~\ref{fig:1}), that is situated beneath the spine of the prominence, have been investigated. The bubble is seen in the form of a dark cavity in the H$\alpha$ line center whereas the same appears bright surrounded by dark threads in the EUV observations. From the NVST H$\alpha$ images, it is evident that the cavity region has a distinctively sharp boundary (Figure~\ref{fig:1}c), a typical feature attributed to the prominence bubbles. Further, the bubble interior is composed of fine structures and appears to be filled partially with the material of relatively lesser brightness than the prominence itself. The prominence structure corresponds to a typical ``hedgerow" shape, suggesting that the prominence main body spans obliquely with respect to the line-of-sight \citep{Ryutova2010}. This can be further confirmed by the H$\alpha$ images of the prominence acquired on the subsequent days where the prominence is visible in absorption against the bright disk and spans along the north-south direction (Figure~\ref{fig:1}d--e). Such configuration allows a clear view of the prominence material as the effect of the sky-plane projection of the background and foreground activities remain minimal \citep{Berger2010}. As follows we characterize the small-scale mass motion within the prominence bubble.

\subsection{Morphological evolution of the prominence bubble in H$\alpha$}\label{sec:h-alpha_bubble}
\begin{figure}[!htbp]
\centering
\includegraphics[width=0.82\textwidth, angle=0]{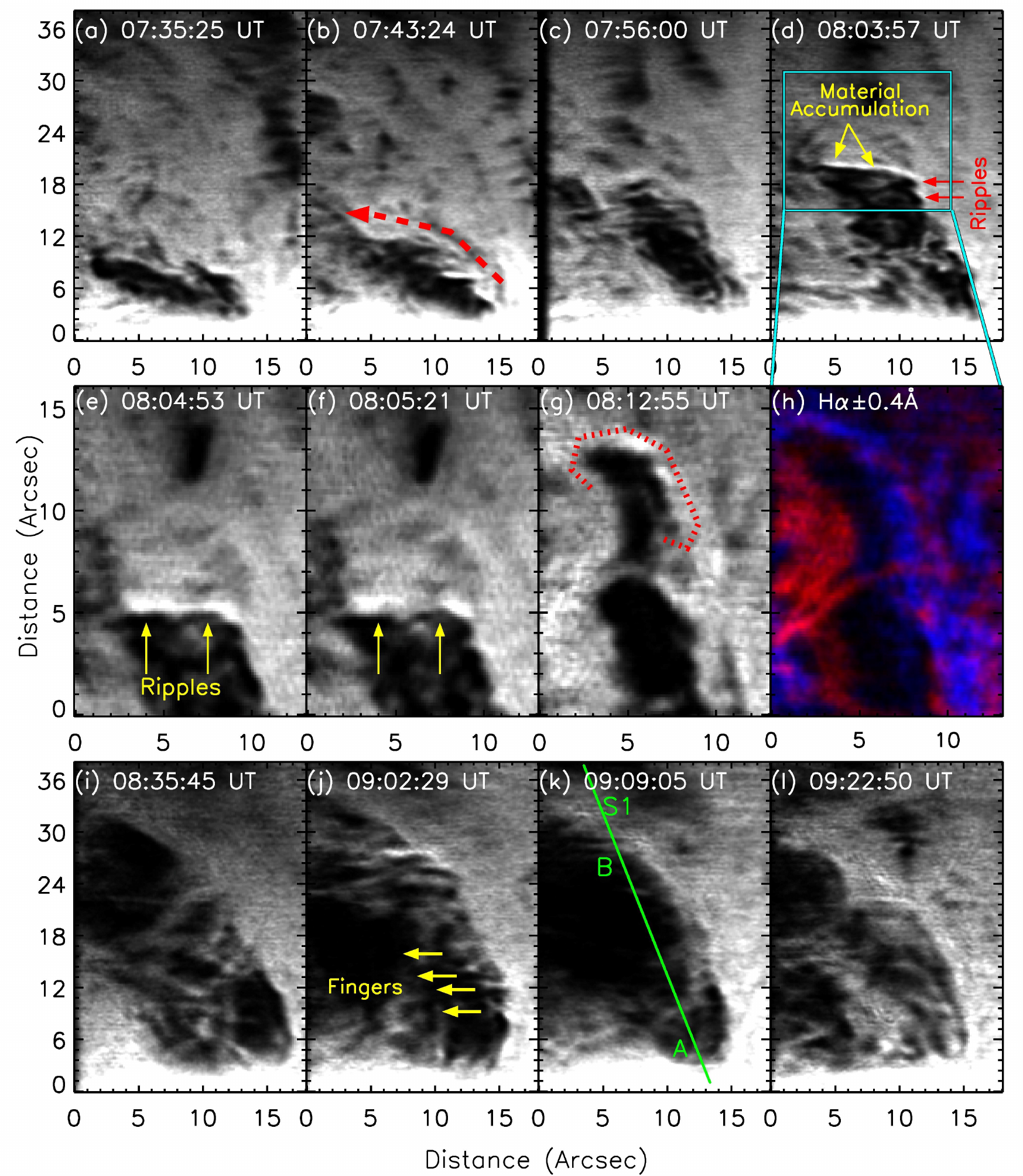}
\caption{Dynamical evolution of a prominence bubble since its formation as seen in the H$\alpha$ line-center and line-wing image sequence, acquired by \emph{NVST} telescope. Crucial dynamical activities exhibited by the material flowing in the bubble interior and at the boundary include; anti-clockwise shear flow at the bubble boundary (panel b), brightened section of the bubble boundary indicating the plasma accumulation as bubble uplifts (panel c); rippling boundary followed by the generation of a mushroom-head plume (panels e--g in H$\alpha$ line-center). H$\alpha$ line-wing images of the plume show dissimilar line-of-sight flow pattern across the rising plume which comprises of blue-shifted material at the head of the plume while red-shifted material is predominant at its left leg (panel h). A clear instance of the formation of finger-shaped structures, extending out from the right boundary of the bubble, is shown in Panel (j). Constantly altering mass distribution in the bubble interior (panels i--l) is indicative of the highly dynamical nature of the investigated bubble.
\\
(\small A movie covering the entire evolutionary sequence of the bubble in the H$\alpha$ line-center as well as in the line-wings, and in various EUV wavelengths is made available at the URL \url{https://rec.ustc.edu.cn/share/552da0a0-1102-11ea-b3c3-17762a199a89}.)}
\label{fig:2}
\end{figure}

The evolutionary sequence of the prominence bubble since its formation has been investigated using the NVST H$\alpha$ line-center and line-wing images (Figure~\ref{fig:2} and associated movie). Prominence bubble originated in the form of an ellipse-shaped void with its major axis having the span of $\sim$13" ($\sim$9 Mm) and acutely tilted towards the solar limb (Figure~\ref{fig:2}a). After 90 minutes of evolution, the bubble enlarged ($\sim$35" (25 Mm)) and became more vertically arranged (Figure~\ref{fig:2}k). Several interesting features have been identified in the bubble interior and at the boundary in the course of its evolution, discussed as follows and in the section~\ref{sec:instability}.

Since the bubble formation, counter-clockwise shear flows are persistently observed along its boundary (Figure~\ref{fig:2}b). Further, as the bubble uplifts, its visually topmost boundary exhibits excess in emission compared to the prominence brightness (Figure~\ref{fig:2}d). This manifests the accumulation of ambient prominence material on the bubble boundary during its uprising process \citep{Ryutova2010}.

\subsection{Doppler map and flow field in the prominence bubble}
\begin{figure}[!htbp]
\centering
\includegraphics[width=0.75\textwidth]{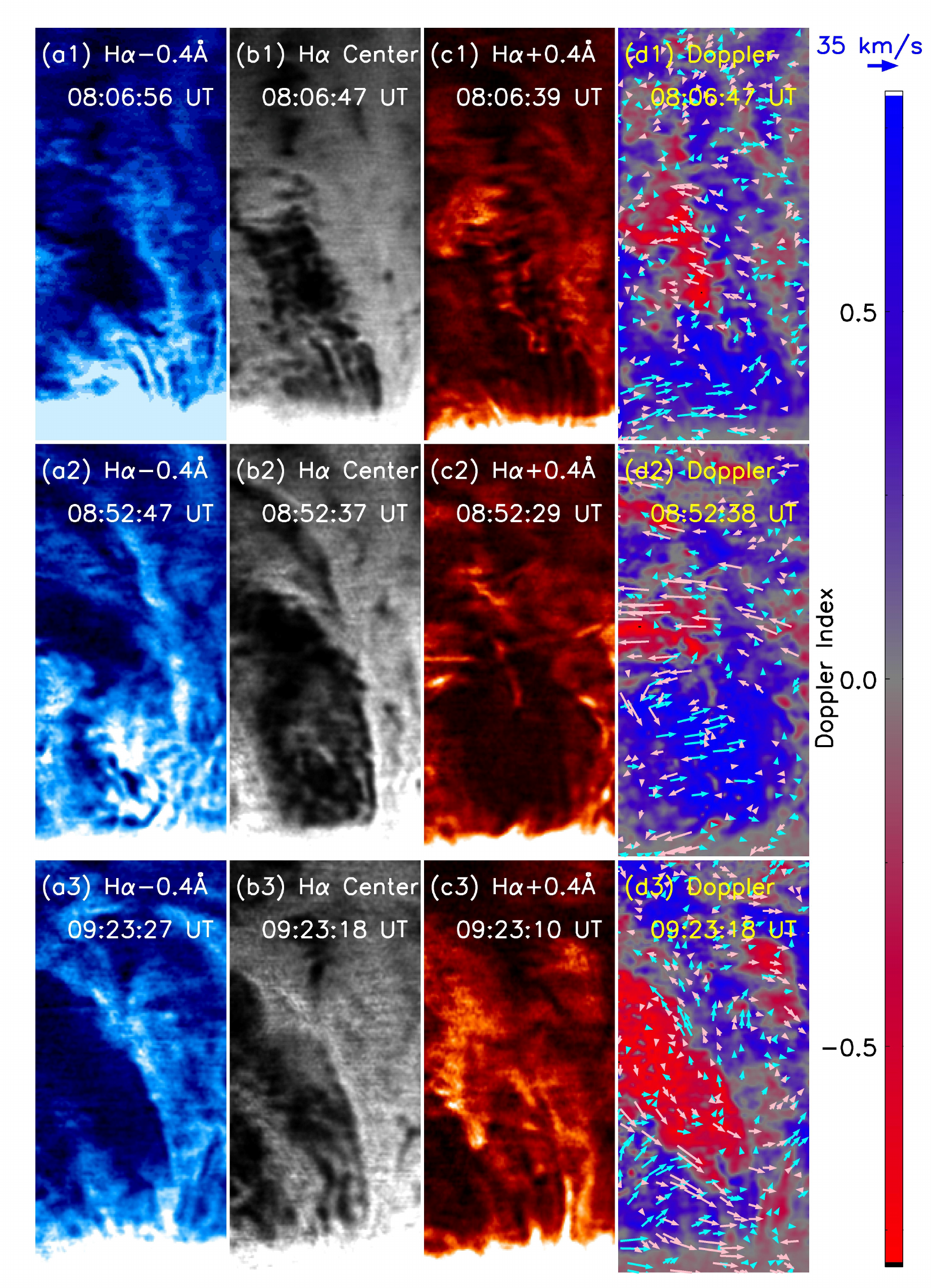}
\caption{Bubble dynamics quantified using the flow maps derived using NAVE procedure and Doppler maps. A sequence of images in the H$\alpha$-0.4 {\AA}, H$\alpha$ line-center, and H$\alpha$+0.4 {\AA} are shown in the first three columns (a--c), respectively. (d1--d3) The flow map derived from the H$\alpha$ blue- and red-wing images is plotted in cyan and pink, respectively, on the respective Doppler maps. The blue (red) color in the doppler map represents the mass motion towards (away from) the line-of-sight. The LOS refers to the direction pointing away and orthogonal to the image plane.}\label{fig:3}
\end{figure}

We investigate the dynamical characteristics of the prominence material by analyzing the images acquired by NVST in the H$\alpha$ line center as well as in the H$\alpha$$\pm$0.4~\AA~wings. It is interesting to note that mass distribution in the bubble interior as seen in the H$\alpha$ line-center images differs remarkably from that appearing in the respective H$\alpha$ blue and red wing images (Figure~\ref{fig:3}). For instance, while the prominence bubble is imaged in the form of a cavity in the H$\alpha$ line-center wavelength at 08:52:37 UT (Figure~\ref{fig:3}b2), the co-temporal H$\alpha$ line-wing images (Figure~\ref{fig:3}a2\&c2) indicate that the mass motion inside the bubble along the line-of-sight (LOS) has a complex pattern. Therefore, multi-wavelength observations are crucial in making a comprehensive assessment of the dynamical characteristics of prominence bubbles, as conducted in the present study.

In order to quantify the mass motion inside the bubble as well as on its boundaries, we employ the nonlinear affine velocity estimator (NAVE) technique \citep{Chae2008a} to derive the flow-map from the NVST images acquired in the H$\alpha$ line-center, blue and red-wing wavelengths. The flow-map has been derived over a grid of uniform spacing of 5 pixels ($\sim$0.5 Mm) and 50~$\times$~65 pixels span. The continuity equation, a default solver in the NAVE procedure, is used for an FWHM of 30 pixels. An important input to the NAVE procedure is the noise level, which is determined using the relationship $\sigma_d/\sqrt{2}$, where $\sigma_d$ is determined as the standard deviation of the absolute difference of two consecutive images corresponding to a region enclosing a quiet and dark area above the prominence. In parallel, in order to deduce plasma motion along the LOS, doppler maps have been constructed using the following relationship \citep{Langangen2008}.
\begin{equation}\label{eq:doppler}
  D=\frac{B-R}{B+R}
\end{equation}
where B and R refer to the pixel intensities in the H$\alpha$ blue and red-wing images, respectively.

The overlay of flow maps, derived using the H$\alpha$ line-wing images (pink (cyan) vectors in Figure~\ref{fig:3}d1--d3 corresponds to the red (blue) wing of the H$\alpha$ line profile), onto the co-temporal doppler maps (background images in the Figure~\ref{fig:3}d1--d3 with the blue (red) color representing the positive (negative) doppler index) revealed the material inside the bubble to be rotating in a counter-clockwise sense. It is further observed that the red-shifted material is predominately exhibiting a downflow (towards the solar disk), whereas a definitive trend of upward motion was seen in the blue-shifted material (Figure~\ref{fig:3}d3). The rotational motion remained persistent during the entire period of investigation (07:27 -- 09:27 UT), however with a varying speed ranging between 5 and 38 km/s. The flow of the red-shifted material has been mainly constrained either at the top of the bubble or at its left side. Similarly, the blue-shifted material appears to be flowing predominately at the bottom or at the right side of the bubble (directions correspond to the vertically upright view of the bubble as a reference, see Figure~\ref{fig:3}).

\subsection{Plasma instabilities in the prominence bubble}\label{sec:instability}
Mass motion inside the bubble and along its boundary reveals several interesting features associated with the plasma instabilities. For instance, at least two distinctively clear small-amplitude (0.5~Mm) ripples are observed at the right boundary of the bubble (shown by red arrows in Figure~\ref{fig:2}d). Subsequently, the top boundary of the bubble also exhibits a clear signature of rippling motions (Figure~\ref{fig:2}e). While there is no significant increase in the amplitude of the rippling motion until the image acquired at 08:05:21 UT (Figure~\ref{fig:2}f), the perturbations increased significantly later, leading to the generation of a typical mushroom-head plume (Figure~\ref{fig:2}g). Interestingly, only a single episode of the plume generation (average uplift speed $\sim$13~km/s) was observed. This indicates the nonlinear explosive phase of the K--H instability to be the most possible mechanism for the investigated plume \citep{Ryutova2010}. To further understand, we have derived the growth rate of explosive instability using equation~\ref{eq:KH_nonlinear} \citep{Ryutova2010} for our case of plume evolution.
\begin{equation}\label{eq:KH_nonlinear}
   \nu_{explosive}\simeq\alpha\frac{\tilde{\nu}_{ei}}{ln(|W|/|W_0|)}
\end{equation}
where the parameter $\alpha$ is considered equal to unity. $\tilde{\nu}_{ei}$ is the rate of inverse slowing down of the particles due to electron-ion collision. Considering the temperature and density of the particles to be 1~MK and 5$\times$10$^{10}$ cm$^{-3}$, respectively, $\tilde{\nu}_{ei}$ takes a value of 2.7$\times$10$^{-2}$ s$^{-1}$. It has been further shown in \citet{Ryutova2010} that the ratio of energy increased ($|W|/|W_0|$) can be approximated to the ratio of the square of the initial and final perturbation amplitudes, which are determined to be 0.5" and 4.5", respectively, in our case (\emph{cf.} Figure~\ref{fig:2}f \& g). With the aforesaid values and equation~\ref{eq:KH_nonlinear}, the rate of growth of the explosive instability is determined to be 9.34$\times$10$^{-3}$ s$^{-1}$, similar to that deduced in \citet{Ryutova2010}. Another possible mechanism for the plume generation can be the coupled KH--RT instability \citep{Berger2017}, however limited spatial resolution restricts the definitive determination of the growth rate of the ripples in their pre-explosive evolution phase (on or before 08:05:21 UT; see Figure~\ref{fig:2}e--f) which makes it difficult to test this scenario. H$\alpha$ line-wing images ($\pm$0.4~\AA) of the plume reveal the blue-shifted emission to be dominant at the plume head whereas the red-shifted material prevails along its left trail (Figure~\ref{fig:2}i--l). This is a possible indication of generation of sheared flow at the plume boundary as the plume ascends.

\begin{figure}[!htbp]
\centering
\includegraphics[width=0.8\textwidth]{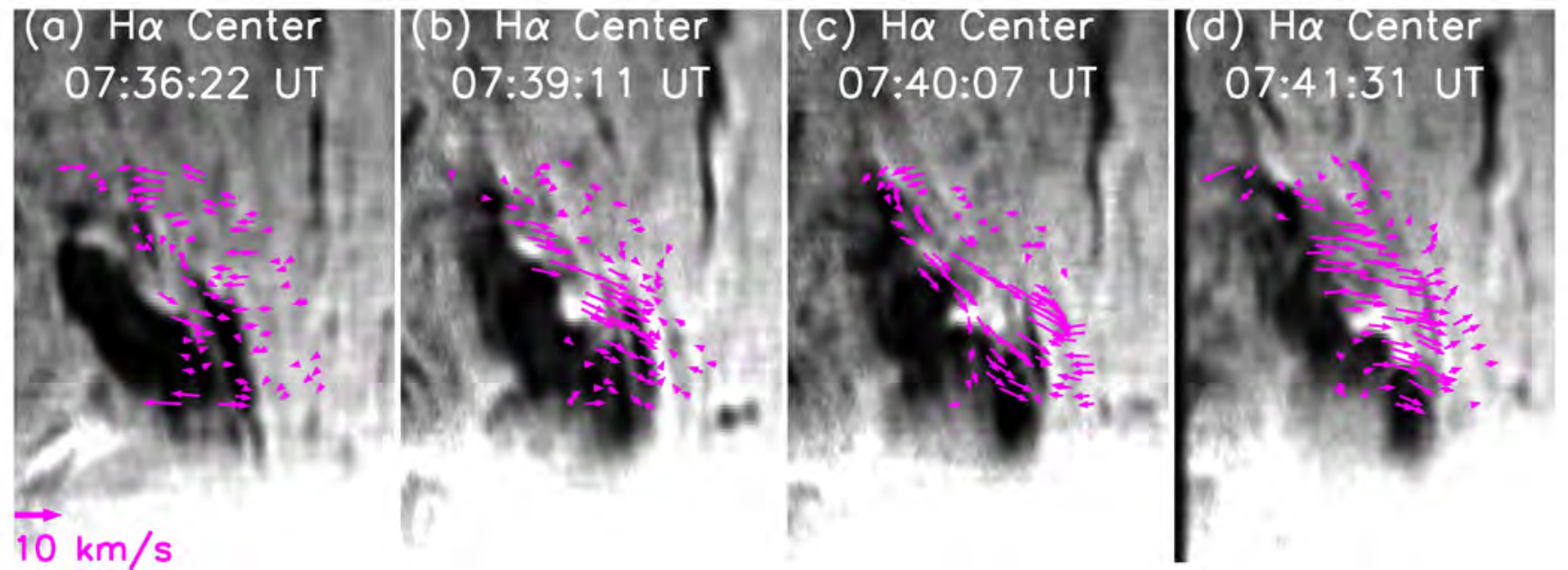}
\caption{Sheared flow along bubble boundary as seen in the early evolution phase of the bubble. Sequence of images in the H$\alpha$ line-center along with the flow-map, derived using NAVE procedure, show the oppositely directed flows along the right boundary of the bubble.} \label{fig:4}
\end{figure}

In addition to the counter-clockwise flow of material, clockwise mass motion along the right boundary of the bubble is also found at the early onset phase of the bubble evolution, particularly during 07:35--07:43 UT (Figure~\ref{fig:4}; also refer to the movie associated with the Figure~\ref{fig:2}). These oppositely directed flows may provide favorable conditions for the generation of K--H instability. Using the NAVE technique on the H$\alpha$ images, we determined the flow speed along the right boundary region of the bubble to be varying in the range of 2-10 km/s, in agreement to that deduced in \citet{Berger2017}. Further definitive characterization of K-H instability may be difficult here due to limited spatial resolution of the NVST images.

Another interesting feature of the bubble evolution is the development of finger structures at the bubble boundary. One such evidently clear instance is reported in Figure~\ref{fig:2}j where the average separation between the fingers is 1.1~Mm ($\sim$1.5"). Usually, finger-like break-up structures are believed to be generated due to R--T instability taking place at the boundary of plasma layers of different densities \citep{Innes2012}. However, since the fingers are not oriented along the direction of solar gravity, these extrusions may be the K--H vortices generated due to shearing flows.

\begin{figure}[!htbp]
\centering
\includegraphics[width=0.8\textwidth]{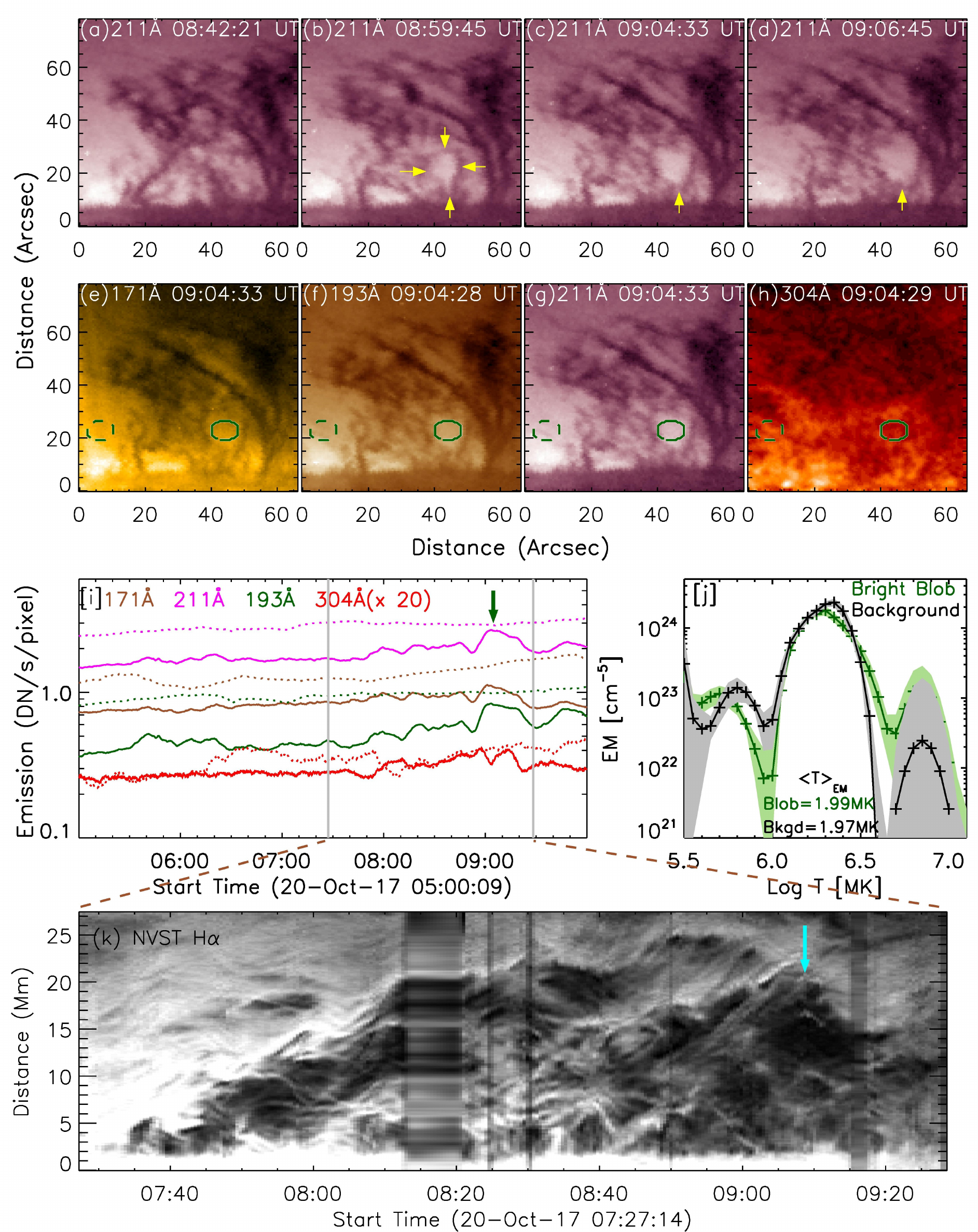}
\caption{EUV perspective of the prominence bubble and thermodynamical evolution of a bright blob within the bubble. (a--d) Similar to the H$\alpha$ observations, the bubble interior appears highly structured throughout its evolution. The formation and expansion of an interesting bright blob-like feature within the bubble, indicated by yellow color arrows in the top panel of the figure. (e--h) EUV images showing the multi-wavelength perspective of the blob at 09:04 UT. (i) Mean EUV intensity profile within a region-of-interest (ROI) representing the blob (green; full line) during 05:00 UT -- 10:00 UT. Background intensity evolution in respective wavelengths as derived from a ROI away from the prominence (dotted green region in panels (e--h)) is also plotted. (j) Emission measure distribution within the selected ROIs corresponding to bright blob (green) and background (black) is plotted along with the respective errors. (k) Time-distance map prepared from the H$\alpha$ line-center image sequence over a virtual slit `S1' along the direction A to B  (see Figure~\ref{fig:2}k).}
\label{fig:5}
\end{figure}

\subsection{EUV perspective and thermal diagnostics of the prominence bubble and bright blob}
Extreme Ultraviolet (EUV) images obtained from the \emph{SDO}/AIA instrument have been analyzed in order to determine the morphological and thermodynamic evolution of the prominence bubble (Figure~\ref{fig:5}). Prominence bubble interior in the 211~\AA~EUV image sequence  (Figure~\ref{fig:5}a; also refer to the movie associated with the Figure~\ref{fig:2}) appears to be highly structured and dynamic in nature, similar to that observed in the H$\alpha$ images.

A distinctively clear bright blob-like feature appeared inside the bubble at 08:59 UT in the EUV images (Figure~\ref{fig:5}b). The evolution of EUV emission corresponding to the bubble is quantified by taking an average of the emission from a small circular region-of-interest (ROI), selected so as to cover the change in intensity from both the bubble interior as well as the bright blob. Similarly, the respective background fluctuations are estimated by averaging the brightness of the pixels corresponding to an ROI away from the prominence, but the same in terms of geometrical parameters (area and radial distance from the limb) of the bubble ROI. The resulted EUV intensity profiles from both the bubble (full lines) and the background (dotted lines) are plotted in Figure~\ref{fig:5}i. We also prepare a time-distance map from the H$\alpha$ image sequence (Figure~\ref{fig:5}k) along a virtual slit crossing the bubble (slit `S1' is shown in the Figure~\ref{fig:2}k) in order to compare the bubble evolution in the EUV and H$\alpha$ wavelengths. Since the earliest formation of the bubble in H$\alpha$, a slight enhancement in the EUV emission compared to that at earlier times is evidenced. In addition to several small-scale perturbations, EUV intensity profile also exhibits a `maximum' at 09:04 UT, corresponding to the bright blob within the bubble. It is crucial to note that the EUV emission corresponding to the blob is always slightly lower than the background emission in the respective wavelengths. This is indicative of the presence of material either in the blob or along its line of sight. On the other hand, the blob appears dark in the H$\alpha$ wavelengths as also seen from the time-distance images during 09:00--09:15 UT (indicated in Figure~\ref{fig:5}k).

To characterize the thermodynamical nature of the blob, we prepared emission measure (EM) maps by employing the method presented in \citet{Su2018}, which is a modified version of the sparse inversion technique for thermal diagnostics developed by \citet{Cheung2015}. This technique makes use of the pixel intensity from the six EUV wavelengths obtained from \emph{SDO}/AIA namely 94~\AA, 131~\AA, 171~\AA, 193~\AA, 211~\AA, and 335~\AA~to derive the EM[T] distribution. We have prepared the EM maps corresponding to the temperature range 0.5 -- 15 MK with a bin size LogT = 0.05. The evolution of emission measure derived by taking an average of the estimated EM[T] distribution over the region corresponding to the bubble and the background ROI are plotted in Figure~\ref{fig:5}j. The uncertainty is calculated using the Monte-Carlo method, implemented in the EM diagnostics technique of \citet{Su2018}. We find the EM values only within the temperature range logT = 6.0--6.6 to be reliable as the derived EM for temperature values beyond this range possess very large uncertainties (Figure~\ref{fig:5}j). Therefore, we estimate the EM-weighted average temperature ($<T>_{EM}$) using the Equation~\ref{eq:t_em} \citep[adopted from][]{Cheng2012} only within the aforesaid temperature range.
\begin{equation}\label{eq:t_em}
  <T>_{EM}=\frac{\sum EM\times T}{\sum EM}
\end{equation}
The $<T>_{EM}$ of the blob results to be 1.99 MK, similar to that of the background corona (1.97 MK). Therefore, this analysis remains inconclusive in untangling the thermodynamical characteristics of bright blob within the bubble from that of the foreground/background corona.

\section{Discussion and Conclusion} \label{sec:conc}
Our investigation of mass motion of a prominence targets the evolutionary phase of a bubble since its earliest appearance in the H$\alpha$ and EUV observations. We find several new morphological and dynamical characteristics of the prominence bubble as discussed following.

\begin{figure}[!htbp]
\begin{tabular}{c}
    \includegraphics[width=0.6\textwidth]{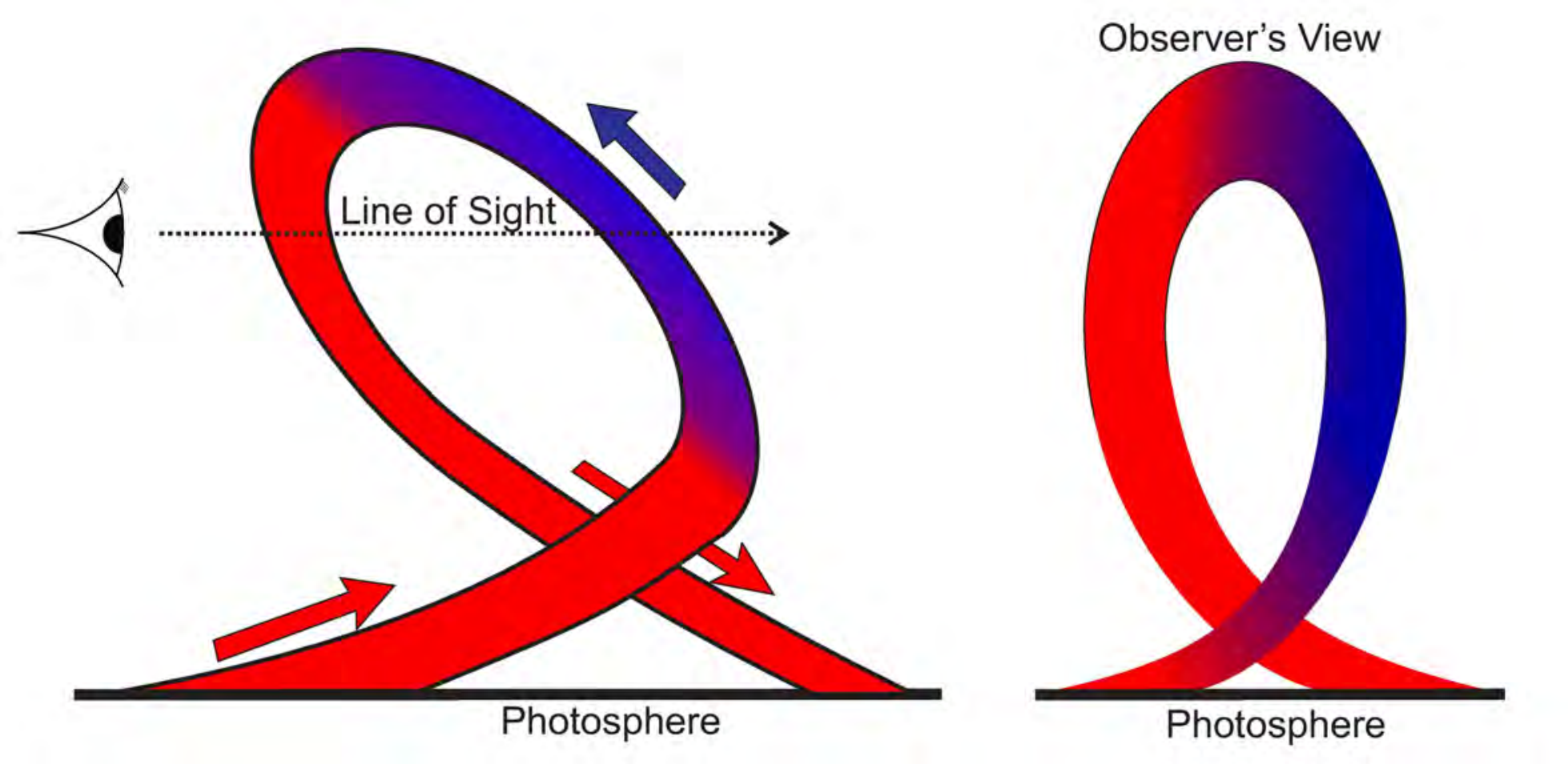}
    \includegraphics[width=0.45\textwidth]{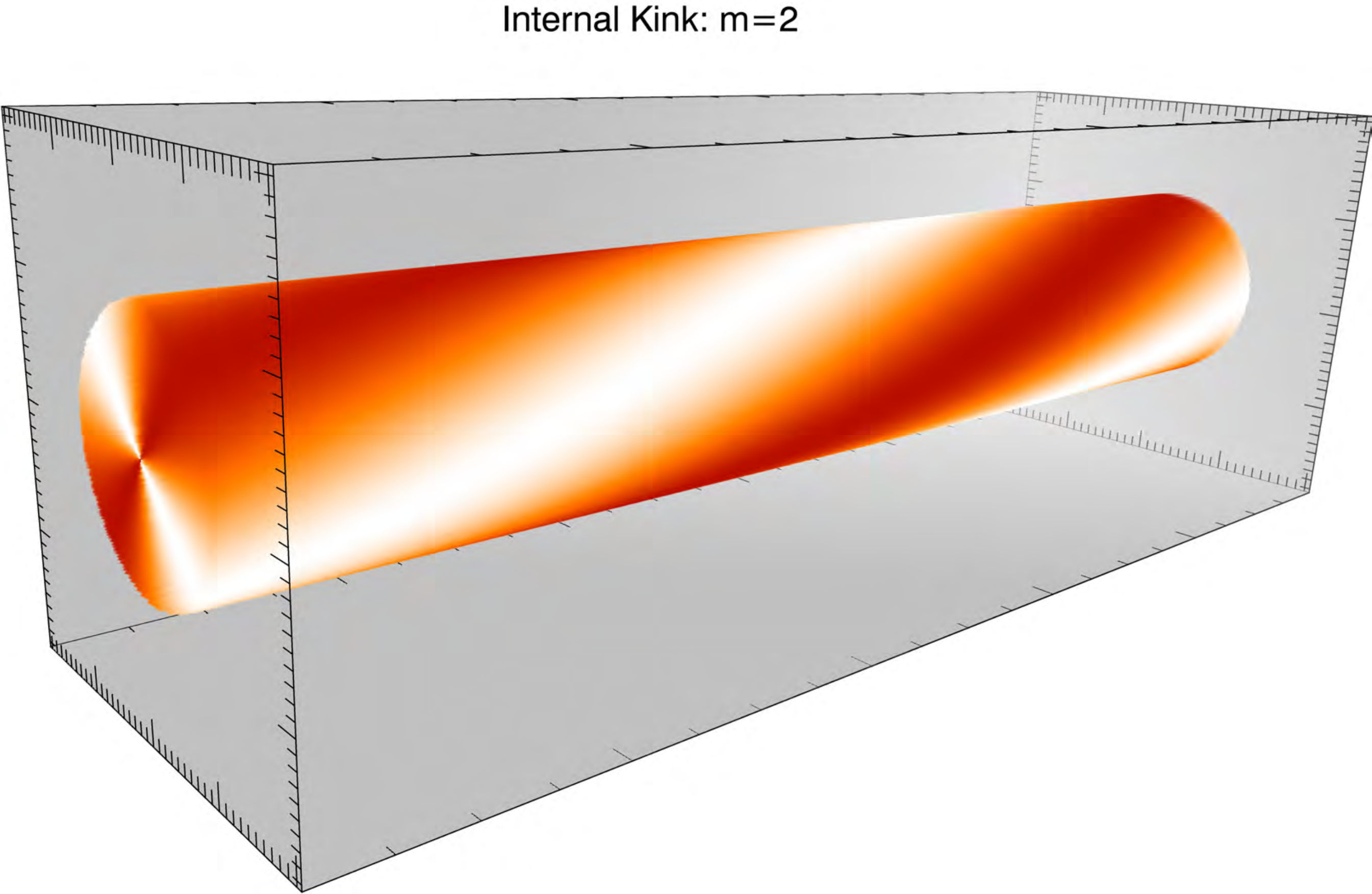}
  \put (-505, 138) {$\textbf{\textcolor{black}{\small[a1]}}$}
  \put (-340, 138) {$\textbf{\textcolor{black}{\small[a2]}}$}
  \put (-161, 140) {$\textbf{\textcolor{black}{\small[b]}}$}
\end{tabular}
\caption{Inferences on the magnetic structure of the prominence bubble as revealed by mass motions. (a1--a2) Schematic representation of a kinked flux-rope as the magnetic field configuration of the bubble, drawn from two perspectives. Material flowing towards (away from) the observer is denoted in blue (red) color. (b) Internal kink instability in a cylindrical flux-rope corresponding to mode (m)=2.}\label{fig:6}
\end{figure}

The bubble interior is observed to replete with dynamic mass (Figure~\ref{fig:2} and Figure~\ref{fig:3}). This suggests that during the formation stage, prominence bubbles do not always possess an obvious cavity-like morphology as usually identified in the existing literature \citep{Dudik2012, Berger2012}. Besides, we have been unable to identify any distinct morphological difference between the bubble location and ambient prominence prior to the formation of the bubble. Therefore, the formation mechanism of the bubble may not require any preferential magnetic field configuration of the pre-existing prominence. The observed disparate mass distribution in the H$\alpha$ line-center compared to that in the co-temporal line-wing ($\pm$0.4~\AA) images indicate a highly dynamical nature of the mass motions inside the bubble (Figure~\ref{fig:3}). To better understand, a comprehensive dynamical characteristic of the prominence bubble is derived by preparing doppler maps and flow-maps from the line-wing images. This revealed a counter-clockwise rotational motion of the material in the bubble interior, which is composed predominately of the blue-shifted material exhibiting upward flow while red-shifted material undergoes a downward flow. Doppler maps further reveal that the red-shifted material is primarily observed in the top as well as at the left portion of the bubble whereas the bottom and the right sections of the bubble are filled with the blue-shifted material (Figure~\ref{fig:3}d1--d3). We interpret this sequence of mass motion to be outlining a kinked flux rope configuration of the magnetic field inside the prominence bubble (Figure~\ref{fig:6}a1--a2). \citet{Liu2007} obtained a similar doppler-shift pattern in a pre-eruptive active-region prominence and inferred it as the signature of a kink-unstable configuration. This concurs with the hypothesis of an emerging flux complex to be the magnetic field structure of the prominence bubble, conceived in \citet{Berger2017}.

\emph{Internal kink instability} \citep{Mikic1990, Hood2009, Keppens2019} can provide an alternate interpretation of the dynamical characteristics exhibited by the prominence bubble investigated in this work. From the counter-streaming mass motions \citep{Zirker1998} in a magnetic field configuration which is resulted from the internal kink instability with higher mode values (m$\ge$2) \citep[refer to Figure~\ref{fig:6}b]{Cap1976}, it is possible to envisage a similar Doppler pattern that is shown by the prominence bubble (Figure~\ref{fig:3}d1--d3). \citet{Mei2018} performed isothermal numerical magneto-hydrodynamic (MHD) simulations in a finite plasma-$\beta$ environment to parameterize the role of internal and external kink instabilities in a magnetic flux rope (MFR). They found both kinds of instabilities to be competing to drive a complex evolution of MFR through the process of reconnection within and around the MFR. However, since the internal kinks are known to possess a smaller growth rate and tend to be energetically benign, which may explain the absence of any obvious signs of heating within the bubble, it can be a preferred mechanism compared to the external kink in the case of prominence bubbles. Further, internal kinks are local and confined in nature, hence their impact on the external field is limited which can help the bubbles in maintaining their shape and boundary for a longer period of time.

Since the earliest appearance of the bubble, we find the signatures of rapid rotational motion within the bubble with a speed much faster relative to the intrinsic motions exhibited by the prominence material. These flows are found to be present within the bubble (Figure~\ref{fig:2}) as well as along its boundary (Figure~\ref{fig:4}) and can be characterized as shear flows \citep{Berger2017}. During the uprise and expansion phase of the bubble, prominence material gets accumulated on the boundary of the bubble. When the shear flow interacts with these dense bubble boundaries, ripples of 0.5-1 Mm amplitude are generated. The amplitude of the ripples rapidly increases in time, leading to generating a large typical mushroom-headed plume. While the ripples are understood to be the signatures of linear phase of instability, its rapid growth rate leading to the generation of a plume is attributed to the non-linear explosive stage of K--H instability \citep{Ryutova2010}. In addition, finger-shaped structures are also observed on the bubble boundary. Although such structures are generally associated with the R--T instability \citep{Innes2012}, we believe that these extrusions may be the K--H vortices as the fingers are not oriented along the direction of solar gravity. Therefore, the generation of K--H instability can be understood as the intrinsic dynamical characteristic of the prominence bubble during its evolution and expansion.

In order to probe the signatures of heating within the bubble, we estimate the EUV emission originating within the bubble region. During the bubble expansion, slight increase in the EUV emission is found inside the bubble (particularly in the 171~\AA, 193~\AA, 211~\AA, and 304~\AA~wavelengths), but the peak emission in the respective wavelengths has always remained lower than that resulting from the background corona (Figure~\ref{fig:5}i). An in-depth investigation further revealed an interesting episode of the formation of a localized blob within the bubble, which appears bright in all of the aforementioned EUV wavelength channels. Similar EUV emission characteristics have been exhibited by a compact region within the prominence bubble investigated in \citet{Berger2011}, who derived its temperature to be of the order of 1~MK. In agreement, the EM-weighted mean temperature of the blob in our case is derived to be$\sim$1.99 MK. However, since ambient corona is also estimated to have similar temperature as that of the blob, it is difficult to infer whether the emission corresponds to the `hot compact region' within the bubble \citep{Berger2011} or from the foreground/background prominence-corona-transition-region (PCTR) \citep{Gunar2014}. Intriguingly, the blob is observed to push the material towards the bubble boundaries during the course of its evolution, which appears to result in the upward expansion of the bubble.

To conclude, high-resolution observation of the prominence not only offers insights into its magnetic field configuration, but it also provides a platform to characterize the generation and growth of the instabilities in the magnetohydrodynamic fluids. Intrinsic mass motions in the prominence (not necessarily leading to its eruption) are an outstanding indirect probe of the physical conditions, as demonstrated in this study where they outline a kinked flux rope configuration of the prominence bubble or provide a new observational signature of the internal kink instability in the prominence. This work also provides physical constraints in the form of morphological characteristics, growth rate, and thermodynamical characteristics of the bubble, which can be used to drive realistic numerical simulations.

\section*{Conflict of Interest Statement}
The authors declare that the research was conducted in the absence of any commercial or financial relationships that could be construed as a potential conflict of interest.

\section*{Author Contributions}
Arun Kumar Awasthi conducted the data analysis and wrote the manuscript under the guidance of Rui Liu. Rui Liu led the interpretation of the results.

\section*{Funding}
A.K.A. acknowledges the support from the Chinese Academy of Science (CAS) as well as the International Postdoctoral Program of the University of Science and Technology of China. R.L. acknowledges the support from NSFC 41474151, 41774150, and 41761134088.

\section*{Acknowledgments}
This investigation primarily made use of the data acquired from the New Vacuum Solar Telescope under a guest observation program. NVST is operated by the Yunnan Astronomical Observatory, Kunming, China. AKA acknowledges the hospitality offered by the staff of Fuxian Lake Solar Observatory during his stay for the observing period and for carrying out the post-processing of the raw data in terms of the alignment and application of speckle-reconstruction technique. RL thank Prof. Jongchul Chae for providing the NAVE code. Dr. Yang Su and Dr. Mark Cheung are acknowledged for the code used to derive the thermodynamical properties of the plasma bubble from the EUV images. Authors also acknowledge the reviewers for their constructive comments which improved the scientific clarity of the work.

\section*{Supplemental Data}
Associated with Figure 2, a supplementary movie covering the entire evolutionary sequence of the bubble in the H$\alpha$ line-center as well as in the line-wings, and in various EUV wavelengths is made available online.

\section*{Data Availability Statement}
H$\alpha$ observations from NVST, analyzed in this work, can be requested through the URL \url{http://fso.ynao.ac.cn/datashow.aspx?id=1782}. Data from \textit{SDO} and \emph{GONG} H$\alpha$ network are archived at the respective instruments' URLs and freely available for download.


\end{document}